# The merging history of the Milky Way

M. Unavane,[1] Rosemary F. G. Wyse[1,2,3] and Gerard Gilmore[1]
[1] *Institute of Astronomy, University of Cambridge, Madingley Road, Cambridge, CB3 0HA*
[2] *Department of Physics and Astronomy, Johns Hopkins University, Baltimore, MD 21218, USA*
[3] *Center for Particle Astrophysics, University of California, Berkeley, CA 94720, USA*

**ABSTRACT**

The age distribution, and chemical elemental abundances, of stars in the halo of the Milky Way provide constraints on theories of galaxy formation. As one specific example, the accretion of satellite galaxies similar to the present retinue of dwarf spheroidals (dSphs) would provide an observable metal-poor, intermediate-age population. This paper presents a quantitative assessment of any contribution made by such stars to the stellar halo. The bulk of the stellar populations in the halo show a well-defined turn-off, at B−V∼ 0.4, implying that the vast majority of the stars are old. The fraction of stars which lie blueward of this well-defined turn-off, with metallicities similar to that of the present dSphs, is used in this paper to place limits on the importance of the recent accretion of such systems. Very few ( $\lesssim$ 10%) stars are found to be bluer (and by implication, younger) than this limit, with the highest value found for the more metal-rich halo ([Fe/H] $\gtrsim$ − 1.5 dex). Direct comparison of this statistic with the colour distribution of the turnoff stars in the Carina dwarf allows us to derive an upper limit on the number of mergers of such satellite galaxies into the halo of the Milky Way. This upper limit is ∼ 40 Carina-like galaxies. The higher metallicity data constrain satellite galaxies like the Fornax dwarf; only $\lesssim$ 5 of these could have been accreted within the last $\lesssim$ 10 Gyr. We note that the low star-formation rates inferred for dSphs predict distinctive elemental abundance signatures; future data for field halo stars, including candidate younger stars, will provide a further robust test of accretion models.

## 1 INTRODUCTION

Observational evidence has been accumulating that merging of smaller systems has played a role in the evolution of the Milky Way and other galaxies. For example, age spreads in globular clusters have been derived (e.g. Searle 1977; Zinn 1993), kinematic 'moving groups' have been tentatively identified in the Galactic halo (e.g. Arnold & Gilmore 1992; Majewski 1993) and disturbed morphology suggestive of interactions and merging is seen in many high redshift galaxies (Griffiths et al. 1994). This is at least qualitatively consistent with hierarchical-clustering scenarios for the growth of structure in the Universe (e.g. review by Silk & Wyse 1993). In those theories the masses of the first objects to turnaround and collapse under self-gravity are approximately equal to those of present-day dwarf galaxies, which then cluster to form larger-scale structure, with this process likely continuing on galactic scales today. The star formation histories in these 'building-blocks' can be very varied, and at present cannot be predicted with any level of certainty by theory.

Quantification of the importance of merging is complicated by the fact that even for purely dissipationless systems (to which we restrict the discussion in this paper), the many parameters that are needed to describe a merger event give rise to many different outcomes. Thus one must be rather specific about the type of merger. Accretion of a dense, fairly massive satellite onto a pre-existing thin stellar disk can inflict significant damage to the disk, during the many passages through the disk that the satellite makes (e.g Toth & Ostriker 1992). Quinn, Hernquist & Fullagar (1993) find that accretion of a stellar satellite galaxy with 10% of the mass of the disk, and mean density within its half-mass radius equal to fully 75% of the central density of the disk, can produce a thick disk with scale-height similar to that of the Galactic thick disk. However, this is a rather extreme satellite, and less robust galaxies are more likely to be tidally disrupted early in the accretion process (e.g. Ostriker & Tremaine 1975).

The discovery of the Sagittarius dwarf spheroidal galaxy (Ibata, Gilmore & Irwin 1994), apparently in the process of being digested by the Milky Way, argues quite irrefutably for on-going accretion, albeit of fluffy objects. Further, the Small Magellanic Cloud appears to be significantly distended along the line of sight, suggestive of strong tidal interaction with the Milky Way. The Magellanic Stream is plausibly also a tidal effect (Lin & Lynden-Bell 1977). Stars which become unbound from the outer regions of satellite galaxies, beyond the tidal radius set by the Milky Way gravitational potential, will remain on orbits close to that of the satellite galaxy at the time of their evaporation. Thus, provided that the satellite is tidally disrupted prior to significant orbital decay through dynamical friction, shredded stars may be expected to contribute to the stellar halo. Probably all of the satellite galaxies to the Milky Way contain a significant intermediate-age ($\lesssim 10$ Gyr) metal-poor population (e.g. Aaronson 1986); as we discuss below, this provides a distinctive signature of their contribution to the field halo.

Globular clusters are a traditional tracer of the stellar halo, even though they amount to only a few percent of the luminous mass. The Sagittarius dwarf spheroidal shares with the Fornax dwarf spheroidal the characteristic of having its own system of globular clusters (e.g. Ibata et al. 1994; Da Costa & Armandroff 1995). The search for Galactic halo globular clusters which are significantly younger than the dominant population, and which could have been accreted from a parent dwarf galaxy with an intermediate-age stellar population, is producing several strong candidates in addition to these Sagittarius clusters (e.g. Fusi Pecci et al. 1995) We here will focus on the field halo itself.

One of the defining characteristics of the field population II is its clear lack of massive main sequence stars (Lindblad 1926; Bok & McRae 1941; Baade 1944; Elvius 1962; Upgren 1963; Sandage 1969). The distinct turn-off of the stellar halo, at $B-V \simeq 0.4$, is a striking feature of deep star counts (e.g. Gilmore & Wyse 1987, their Fig. 3). This turn-off colour corresponds to an age, derived from comparison with stellar isochrones, of $\gtrsim 15$ Gyr, for a population with the mean metallicity of the stellar halo, $<$[Fe/H]$>\sim -1.6$ dex. Thus, any halo stars significantly younger than this are restricted to a tracer population.

However, it has been realised for many years that such a tracer population may exist. A few high-velocity dwarf stars bluer than the halo turn-off were identified in early surveys (e.g. Bond & MacConnell 1971).[*] Preston, Beers & Schectman (1994) analysed their sample of stars, selected on the basis of the weakness of their Calcium H and K lines, and concluded that a significant fraction were dwarf stars bluer than the halo turnoff. The absolute normalisation is difficult to quantify given the selection criteria of the sample, but they suggest that perhaps 4%–10% of the stellar halo is in this component, which they attribute to the accretion of dwarf spheroidal satellites.

Quantification of the fraction of the stellar halo that is younger than the dominant population requires careful analysis of samples where the selection effects are understood. We here discuss the distribution of main sequence field halo stars in the [Fe/H]–(B−V) plane, exploiting the contrast between the stellar population of a typical dSph – a metal-weak, [Fe/H] $\lesssim -1$ dex, intermediate-age $\lesssim 8$ Gyr, population – and a background, old, halo population.

More qualitative limits only can be placed at present using other tracers of intermediate-age population, such as carbon stars. We make predictions for the distinctive element ratios that are possible for stars in the younger populations of dSph – and thus stars from this population accreted into the halo.

## 2   THE AGES OF THE HALO

Searches for tracer populations obviously work best when focussed on large, representative samples. The low local normalisation of the field halo has led naturally to two main approaches to the study of the halo, each of which has its advantages and disadvantages – (i) identification of halo stars by virtue of their current large distance from the disk plane and (ii) identification of local halo stars by their high velocities with respect to the Sun.

The major disadvantage of the *in situ* approach is that the fraction of the halo studied is in general rather small, and one may be sensitive to local fluctuations, especially in light of the fact that dynamical times in the outer Galaxy are longer. Quantification of the fraction of the halo that lies beyond a given radius and/or height above the plane is given in Fig. 1, which is based upon the Hernquist (1990) density profile. This is an analytic approximation to the deprojection of a de Vaucouleurs profile, and has half of the total mass contained within a (deprojected) radius equal to 1.33 times the (projected) de Vaucouleurs effective radius. As seen in the figure, for the stellar halo, which has de Vaucouleurs effective radius of $\sim 2.75$ kpc, the half-mass radius is $\sim 3.7$ kpc.[†]

Star count data and models suggest that the thick disk makes a significant contribution out to $\gtrsim 5$ kpc above the disk plane, so that a 'pure halo' sample is better isolated beyond this. However, as shown in the figure, even an all-sky survey out to infinity would only sample $\lesssim 10\%$ of the halo, defined this way. A typical pencil-beam survey towards the Galactic Pole, of $\lesssim 1$ deg$^2$, samples only a few times $10^{-6}$ of the halo when restricted to $z > 5$ kpc (e.g. Majewski 1992). Of course one assumes that this represents a fair sample of the distant halo, but its relation to the *total* halo is not so obvious.

Kinematically-selected local samples of high-velocity stars lack those stars with space motions too close to that of the Sun, and also stars on orbits which do not intersect the solar circle. However, solar-neighbourhood surveys of high proper-motion stars provide a reasonable sampling of the halo, with $\gtrsim 30\%$ of the stars in a plausible halo distribution function being observable (e.g. May & Binney 1986).

---

[*] More recently, a very few ($\lesssim 10$ in the whole Galaxy) young main sequence B stars, which apparently have formed *in situ* in the halo, have been identified (Conlon et al. 1992). These B-stars are however very metal-rich, and are unrelated to the halo field population which is the subject of this paper.

[†] By comparison, the half mass radius of an exponential disk is $\sim 1.7$ times the disk scalelength, and for the Milky Way disk this is then $1.7 \times 3 - 4$ kpc, or $5 - 7$ kpc.

**Figure 1.** The mass fraction of the halo which lies (a) beyond galactocentric distance $r$ – upper panel – and (b) outside the galactocentric distance $r$ associated with a height $z$ above the galactic plane at the solar circle ($r_0 = 7.8$kpc) – lower panel. In each case the solid line represents a spherical halo model, and the dotted line represents an oblate halo, with axial ratio of 0.6, both based on Hernquist (1990) density profiles.

**Figure 2.** The distribution of $B-V$ colour and [Fe/H] for metal-poor stars from the proper-motion selected sample of Carney et al. (1994). Uncertainties have been ignored in the interests of clarity, and are of order 0.1 dex in [Fe/H] and 0.05 in $B-V$. Superposed are turn-off isochrones (*Revised Yale Isochrones*) with ages given in Gyr. The vast majority of stars have colours consistent with ages $\gtrsim 15$ Gyr. The three metallicity ranges delineated by the dashed lines are discussed separately.

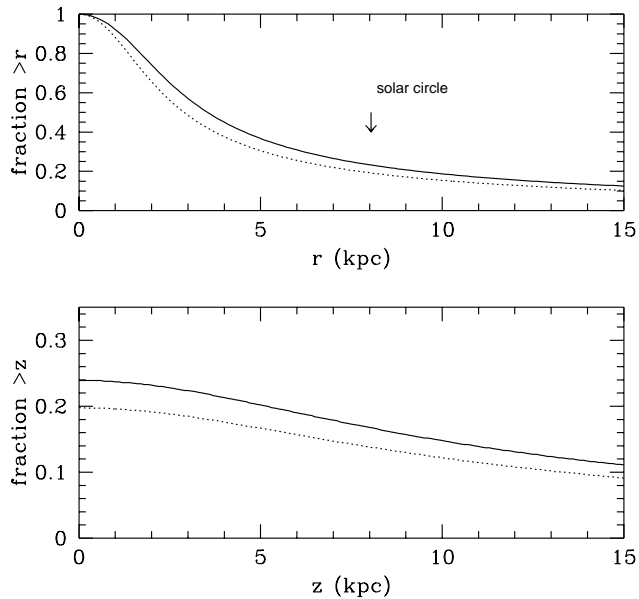
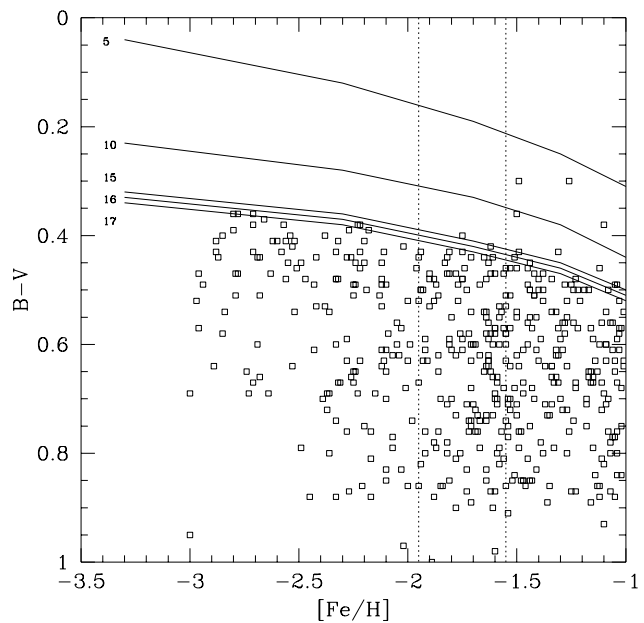

Direct age estimations for individual field halo stars are available only for the samples of Schuster & Nissen (1989) and Marquez & Schuster (1994). These surveys derive ages through comparison of Strömgren photometry and isochrones. While the overall conclusions of these surveys are consistent with the bulk of the halo being old, the uncertainty associated with any individual age determination is still too great for an analysis of the type we now pursue.

The recent survey by Carney et al. (1994 and references therein) contains the largest sample of halo stars (1447 stars), selected on the basis of proper motions, for which accurate abundances and photometry are available. This survey provides reliable $B-V$ photometry and metallicity estimates, in addition to the space motions (radial velocities and proper motions), for all except the reddest stars. Metallicities and colours are given for 1229 of these stars. All stars, including those few in Common Proper Motion pairs, have been used in this investigation. The data of most interest for present purposes are those for the metal-poor halo, which for this sample we will initially define by the condition [Fe/H] $\lesssim -1$. This gives a sample of 477 stars, representing the population of the metal-poor halo. The relationship between $B-V$ colour and [Fe/H] for this sub-sample is indicated in Fig. 2.

The colour distribution in Fig. 2 shows clearly the turnoff of an old, metal-poor population, with the turnoff colour being a function of [Fe/H]. In order to quantify the number of stars which may lie to the blue of this turnoff, we employ isochrones as an objective measure of the colour-dependence of the edge of the distribution. We have superposed the *Revised Yale Isochrones* (Green et al. 1987) interpolated to Y=0.24. The Vandenberg & Bell (1985) and Vandenberg (1985) isochrones lie parallel to these, differing only by a colour offset of about 0.06 in B-V, with the *Yale* isochrones being redder for a given age. We emphasise that the absolute age scale of the isochrones is not relevant here, only their shape is relevant. The $\gtrsim 15$ Gyr isochrones adequately represent the cut-off colour for the majority of the stars; bluer stars are rare. There is clearly room for a couple of Gyr scatter about the age of the isochrone defining the dominant population, especially since the $B-V$ turn-off colours of the old isochrones crowd together. Indeed, a recent determination of the ages of 43 halo globular clusters (Chaboyer et al. 1995) indicates a dispersion in the age of the old, metal-poor globular clusters of about 2 Gyr. Our aim here is to determine a firm *upper limit* to the number of field halo stars that are much younger than the dominant old population, with 'old' encompassing the $\sim 2$ Gyr scatter in age that this population appears to have.

In order to quantify the number of stars which are significantly bluer than the dominant turnoff, we proceed, preempting the discussion of Section 3.1 below, by dividing the metallicity range into three regions: [Fe/H] $< -1.95$, $-1.95 \leq$ [Fe/H] $< -1.55$, $-1.55 \leq$ [Fe/H] $< -1.0$. We illustrate the calculation for the intermediate abundance interval. This sub-sample is indicated on Fig. 2 by vertical dashed lines, and is shown enlarged in Fig. 3. The observational uncertainties in $B-V$ from Carney et al. are indicated, while the uncertainties in [Fe/H], typically of order 0.1 dex, are not indicated, in the interests of clarity of the figure. The slope of the relationship between turn-off colour and metallicity is such as to minimise the effects of these uncertainties on a given star's location relative to the turn-off isochrone. We have chosen the 16 Gyr Yale isochrone as fiducial. It is clear that though two points lie blueward of this isochrone, it is not possible to assign them with certainty to the class of 'younger' stars. Assuming Gaussian errors, there remains a 9% probability that both are *not* too blue, and a 45% probability that one of the two is not too blue. Weighting these possibilities appropriately, and allowing for the errors in slightly redder stars, we deduce an expected value of 2.2 stars out of the 147, or 1.5%, as being significantly bluer than the bulk of the population, in the range $-1.95 \leq [Fe/H] < -1.55$

Assuming now that these stars are indeed younger, and have an associated luminosity function, this number must be multiplied by an appropriate factor to take account of the redder, lower-mass main-sequence stars that would be part of this younger generation. Taking the globular cluster M5 as typical of a metal-poor population (to translate B−V to corresponding absolute magnitudes) the range of colours in Fig. 2 which contains the 147 stars in the metallicity range under study, corresponds to $M_V = +4$ and $M_V \sim +7.5$, while the bluer stars of interest have $M_V \lesssim +4$. The luminosity functions for both globular clusters and for the solar neighbourhood are quite flat over this range, with an approximately equal number of stars in each interval of unit absolute magnitude. Thus we can estimate the number of lower main-sequence stars which should be associated with each blue star by applying a correction factor of $\sim 4$. Thus $\sim (2.2 \times 4)/147$ or $\lesssim 6\%$ of the field halo, in the metallicity range $-1.95 \leq [Fe/H] < -1.55$ plausibly have younger age than the dominant old population.

The bluest stars in Fig. 2 are predominantly more metal-rich than −1.5 dex. The six bluest of these stars were in fact identified in Table 5 of Carney et al. (1994) as 'blue straggler candidates', although they did not comment further on their significance. A similar calculation to that illustrated above, again allowing for the probable errors in the colours of the sample, and assuming them to be Gaussian, provides a total of 13.1 stars (out of 185) expected to lie blueward of the 16Gyr Yale isochrone, in the region $-1.55 \leq [Fe/H] < -1.0$. That is, 7.1% of the 185 stars in that metallicity range, corresponding to 28 percent possible young population, after weighting to allow for the lower mass contribution.

The contribution from the 145 stars with [Fe/H]< −1.95, taking account of the Gaussian errors, gives a total of 1.1 star, or < 1%, bluer than the old isochrone. This corresponds to 3 percent of this metallicity range possibly belonging to a younger population, after correction for the luminosity function.

To summarise, after weighting to take account of the luminosity function, the upper limits on younger star population in the field halo comprises $\sim 3\%$ in the most metal-poor range, $[Fe/H] < -1.95$; $\sim 6\%$ in the middle range, and $\sim 28\%$ in the most metal-rich range considered, $-1.5 \leq [Fe/H] < -1$.

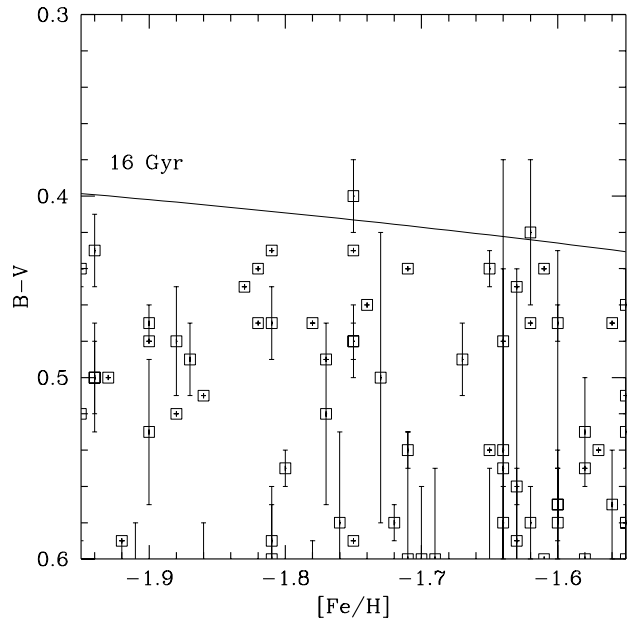

**Figure 3.** The same as Fig. 2, except enlarged for the region $-1.95 \leq [Fe/H] < -1.55$, and with the observed uncertainties in B−V indicated, when available. We can conclude that the expectation is that 2.2 stars in 147, in this range of metallicities, lie blueward of the 16 Gyr Yale isochrone.

### 2.1 Caveats

There are two points of note. Our assumption – that all stars found possibly to the blue of the main-sequence turnoff, after taking account of estimated uncertainties in the photometry, are genuinely young halo stars – is conservative and leads to an upper limit to their contribution. A large fraction of the signal is due to stars which are formally to the red of the 16 Gyr isochrone, but whose photometric errors are such that there is some, albeit small, probability that they actually lie to the blue of the isochrone, and are hence counted using our technique. Further, any metal-poor thick disk stars which might exist, and might be younger than the halo, will be included.[‡] In addition, blue stragglers of the type found in the cores of globular clusters, and mis-identified evolved stars, could all contribute. An estimate of the expected number of such blue straggler stars may be made using the compilation of globular cluster data by Ferraro et al. (1995), under the assumption that the number-to-light ratio in the field halo is the same as that in loose globular clusters. This compilation contains results for 26 globular clusters, from which Ferraro et al. find approximately 7 more blue straggler stars (BSS) for every extra $10^4 L_\odot$ of bolometric intensity sampled, corresponding to $\lesssim 10$ BSS for every $10^4 L_\odot$ of V-light sampled.

---

[‡] The colour–metallicity distribution of thick disk stars is consistent with the bulk of these stars being of the age of 47 Tuc, a typical old, metal-rich globular cluster (e.g. Gilmore & Wyse 1987; Wyse & Gilmore 1988; Gilmore, Wyse & Kuijken 1989; Carney et al. 1989; Gilmore, Wyse & Jones 1995). Age estimates from Strömgren photometry also yield an old age for stars of thick-disk kinematics and metallicity (Edvardsson et al. 1993; Marquez & Schuster 1994).

The V-band luminosity of the field halo is $\sim 10^9 L_\odot$, so that one would expect of order a million globular-cluster BSS in the entire halo. The present halo sample is just dwarf stars of high proper motion. Somewhat less than half, perhaps $\sim 30\%$, of the V-band light in an evolved population comes from the main sequence and subgiant stars (e.g. Tinsley & Gunn 1976); thus one would expect $\lesssim 10$ BSS per $5 \times 10^3 L_{V,\odot}$ from dwarf stars in the halo. The total V-band luminosity of the present sample of $\lesssim 500$ dwarf stars will be around $\lesssim 200 L_\odot$ (as may be confirmed by simple addition of all the estimated individual luminosities). Thus we expect only of order *one* globular-cluster like BSS in our sample. Even taking account of the uncertainties in this rough calculation, it is unlikely that these stars contribute significantly to the blue stars we identify.

More substantially, we have analysed a sample selected by high proper-motion. There is no robust *a priori* expectation that the kinematics of any young halo population should be similar to those of the older halo stars. If the younger stars do result from the accretion of satellite galaxies, their orbits will reflect those of the parent satellite galaxies at the time the stars were captured, and dynamical friction may well have altered the satellite's orbit.

Some general points may be made concerning the effects of selection by proper motion. Should the mean rotational lag of a stellar population be greater than the tangential velocity dispersion of that population, then the probability of selection into the catalog is dominated by the azimuthal streaming velocity (cf. Ryan & Norris 1993). This will be the case for populations with velocity dispersions of $\gtrsim 100$ km/s. Such a stellar population will be efficiently selected by a proper motion criterion irrespective of its velocity dispersion. That is, any stellar population with asymmetric drift greater than $\sim 100$km/s will be found in a proper motion sample almost irrespective of its velocity dispersion.

It should be noted that in general dynamical friction removes both orbital angular momentum and orbital energy, so that a small lag behind the Sun could only be obtained for a population of stars from accreted satellite galaxies if the parent galaxies were originally on exceptionally high angular momentum orbits, with mean azimuthal streaming much higher than that of the field halo. This would require that the satellite galaxies would have exceptionally high peri-Galactic distances. This in turn implies that dynamical friction was relatively inefficient, resulting in long accretion times. For any putative satellite which had actually been accreted, it is more likely that the accreted stars should have a large mean lag in rotational velocity with respect to the Sun, and thus find their way efficiently into a proper motion selected sample of stars near the Sun.

A sample, with kinematic data, which is suitable for comparison here is that of Preston et al. (1994). They determined that their sample of 'Blue Metal-Poor stars' (BMPs), which have metallicity [Fe/H] $\lesssim -1$ dex and which they suggest consists of truly young stars, does have kinematics which distinguish it from the classical halo. They found that the BMPs have a mean rotational streaming velocity of amplitude 128 km/s, while that of the old halo, at the same metallicity, is $\sim 25$ km/s. The BMPs further have an isotropic velocity dispersion tensor, with amplitude $\sigma_{r,\theta,\phi} \sim 90$ km/s, in contrast to the old halo with $(\sigma_r, \sigma_\theta, \sigma_\phi) = (130 : 100 : 90)$ km/s (Gilmore et al. 1989).

Thus the space motion relative to the Sun is (marginally) lower for a typical BMP than for a typical halo star, making it (marginally) less likely that the BMP population is represented fully in a proper-motion-selected sample. However, the kinematics are such that the lag behind the Sun in rotational streaming is likely to dominate any difference in selection probability between old halo and 'young' halo. We note that Preston et al. do not take this effect into account in their calculation of the kinematic biases against discovery of their young stars.

Ryan & Norris (1993) investigated the probability that a star with a given space motion would pass into a proper-motion selected sample of given selection criteria. They calculated the 'survival probabilities' of stars chosen at random from a population of specified mean rotational velocity and velocity dispersion tensor. In situations where the rotational velocity dominates, the survival probabilities, albeit low, do not vary by more than a factor of a few, for rotational velocities varying over the range 100 – 200 km/s behind the Sun. This is comparable to the difference between a typical BMP and a typical old halo star. Thus the under-representation of a BMP in a proper-motion catalogue, relative to the old halo, will likely be only a factor of a few. This is compensated to some extent by our conservative approach in counting stars blueward of the dominant turnoff.

## 3 WHAT FRACTION OF THE YOUNG HALO CAN BE RECENT ACCRETION?

The result above gives the fraction of the halo which may belong to a population that is younger than the majority of the stars. These younger stars may perhaps have been accreted through the tidal disruption of satellite galaxies. The fraction of young stars may then be re-expressed in terms of the number of satellite galaxies that could have been accreted, once one characterizes the stellar populations of the satellite galaxies.

A typical dwarf spheroidal galaxy contains a significant fraction of its stars in intermediate-age or young populations (Hatzidimitriou & Irwin 1995). For example, Mighell (1990) demonstrated that the majority of stars in the Carina dSph were of age $\sim 8$ Gyr, and Lee et al. (1993) have shown that the age of the dominant population in Leo I is rather young, at $\sim 3$ Gyr. Few, if any, have exclusively old populations. Thus, typical dSphs which are accreted recently (after they have formed their younger stars!) into the field halo of the Milky Way will signal themselves by the presence of younger stars. These younger stars manifest themselves in the dSph by a turnoff colour that is bluer than the main sequence turn-off of a $\gtrsim 15$ Gyr population, and by the presence of luminous stars at the tip of the giant branch, perhaps including carbon stars. Such tracers allow a comparison to be made with the stellar halo. To minimise model and isochrone dependences, we make this comparison empirically, adopting Carina as representative of the dSph population.

### 3.1 How Many Carina Dwarfs?

The Carina dSph has recent VR photometry to V$\sim 25$ (Mighell 1990) and BI photometry to B$\sim 25$ (Smecker-Hane et al. 1994; 1995), which suggests that $\sim 80\%$ of the stars (evolved and unevolved) in this galaxy have ages ~ 8 Gyr (Mighell 1990; Smecker-Hane et al. 1994; 1995). The analyses of these colour-magnitude diagrams, together with extant spectroscopy, imply a metallicity of [Fe/H] $\simeq -1.75 \pm 0.20$. The same isochrones that were used above to estimate the dominant turnoff of the field halo, at $\gtrsim 15$ Gyr, provide an old turnoff colour at this metallicity of V−R= 0.26. Thus we can compare the field halo B−V colour distribution to that of the Carina dwarf, by counting stars in the Carina dwarf blueward of V−R= 0.26, selecting stars in the Carina dwarf to isolate the turnoff. The histogram of V−R colours that results from a random subsample of Mighell's (1990) stars fainter than V=22.5 is shown in Fig. 4. The figure clearly demonstrates the existence of a substantial intermediate-age population, with colours bluer than V−R= 0.26, comprising more than one-half of the subsample shown.

This value of 50% may be directly compared with the value for the percentage of the field halo that is blueward of the same isochrone, in this metallicity range, which was 1.5%. Thus 3% of the halo (with this metallicity) has the same colour distribution – and hence inferred age distribution – as the Carina dwarf. Using the Laird et al. (1988) halo metallicity distribution, which is obtained by further selection on their proper-motion selected sample to isolate halo stars, the metallicity range under consideration ($-1.95 \leq$ [Fe/H] $< -1.55$) represents ~ 30% of the entire halo. Thus around 1% of the entire halo could have been formed by accretion of satellite galaxies like the Carina dSph.

The Carina dSph has a total luminosity of $M_V \simeq -9.2$, while the halo has $M_V \simeq -17.5$. Thus this limit of 1% of the halo amounts to $\gtrsim 20$ Carina dwarfs.

The luminosity-function weighted value we obtained above, 6%, for the younger population in the field halo, can be directly compared with the 80% estimate of the total intermediate-age population in the Carina dwarf. This gives a factor of two higher limit on the contribution of Carina-like objects to the halo, i.e. of order 2% of the halo, or $\gtrsim 40$ dwarfs.

### 3.2  If not like the Carina dwarf, then...?

Extending this result to constrain the recent accretion of *all* dwarf galaxies drawn from the present population of satellite galaxies would require detailed colour and metallicity data. Photometric data are not in general available, but reliable metallicity estimates are.

Fig. 5 shows the halo metallicity distribution function taken from Laird et al.(1988), together with the mean metallicities of the seven dSph companions to the Milky Way for which spectroscopic metallicity estimates are available. The composite stellar population of dSph complicates the determination of photometric metallicity estimates from analysis of their color-magnitude diagrams. For example, recent estimates of the photometric metallicity of Leo I range from $\lesssim -2$ dex (Lee et al. 1993) to $\gtrsim -1$ dex (Reid & Mould 1991). We have also not used the published data for the metallicity of Leo II ($-1.9$ dex), which is based on spectroscopy of only three stars (Suntzeff et al. 1986). The upper panel shows these distributions simply normalised to unity in the highest bin. The dwarf spheroidal galaxies have an approximately uniform distribution of metallicity (we return to this point briefly below, in terms of the luminosity function of dwarf galaxies). The locations of the Magellanic Clouds are as indicated. The lower panel compares the *luminosity-weighted* metallicity distribution for the dSphs with the halo metallicity distribution. It is this weighted distribution function which would correspond to the metallicity distribution of stars accreted into the halo from a random subset of the present day dSph luminosity function. It clearly does *not* match the metallicity distribution function of the field halo. While this weighting is rather uncertain, as all dSph parameters are still only approximately known, the luminosity–metallicity relation (e.g. Armandroff et al. 1993) ensures that Fornax and Sagittarius dominate, providing a relatively metal-rich mean to the stars that make up the present dSph population. This weighting, of course, excludes the Magellanic Clouds. They are completely inconsistent with providing any contribution to the field halo. Both the LMC and SMC have metallicities above −1 dex, and are both of comparable luminosity to the halo itself ($M_V = -18.5$ for the LMC, and $M_V = -16.5$ for the SMC, with $M_V = -17.4$ for the halo). The accretion of one of these would lead to a huge contribution to the $> -1$ dex metallicity population in the halo, which is not observed.

**Figure 4.** The distribution of V−R colours for a random subset of stars near turnoff in the Carina dSph. The dotted line indicates the expected turnoff for a population of age 16 Gyr (Yale isochrone). The younger population clearly manifests itself as the $\gtrsim 50\%$ of the stars in this subsample found blueward of this line.

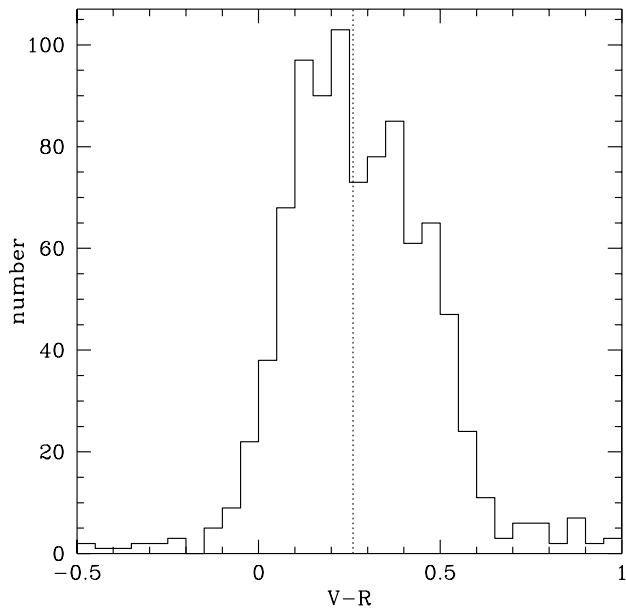

**Figure 5.** Metallicity distribution of seven of the dSph companions to the Milky Way (solid histogram) compared with the field stellar halo (dashed histogram). Only those dSph with mean metallicity derived from spectroscopic estimates from a reasonably large sample of stars have been included. The dSph are represented by their mean metallicity, but one should bear in mind that an internal dispersion of 0.2 – 0.3 dex is typical. The upper panel shows the unweighted distribution, the lower panel shows the dSph weighted by their individual luminosities. The mean metallicities of the Magellanic Clouds are indicated by arrows.

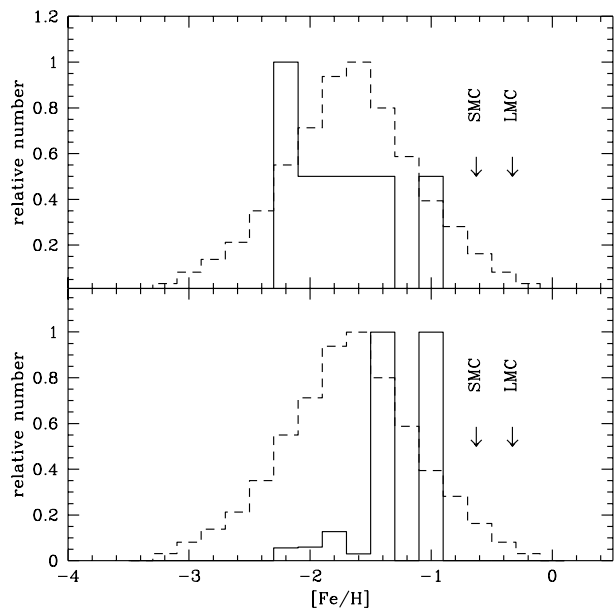

It is clear that the old, dominant population of the halo could not have been formed by accretion of dwarf satellites drawn from a parent population with the same properties as the integrated population of the present-day dwarf spheroidals, or the Magellanic Clouds. Fornax is known to contain a mix of old and intermediate-age stars (Beauchamp et al. 1995), while Sagittarius appears predominantly intermediate-age (Ibata et al. 1994; Mateo et al. 1995). Thus, the signature of recent accretion from the present parent population of dSph would be intermediate-age stars in the halo, with metallicity strongly peaked at [Fe/H] $\gtrsim -1.5$ dex.

This is indeed the metallicity distribution we found for the candidate young halo stars, in section 2 above. Detailed age distributions are lacking for the metal-rich dSph, but one can derive a rough limit on the number of such that could have been assimilated into the halo, by assuming 100% intermediate-age. We found that $\sim 28\%$ of the stars in the Carney et al. sample in the metallicity range $-1.55 \leq$ [Fe/H] $< -1$ could be younger than $\sim 15$ Gyr. Again using the Laird et al. halo metallicity distribution, this range corresponds to $\sim 35\%$ of the halo, so that approximately 10% of the halo is 'young' in this case. The field halo is approximately 50 times more luminous than Fornax, adopting a luminosity of $-13.2$ for Fornax, so that of order 5 systems similar to Fornax are implicated.

This is a small enough number of distinct objects that a signal in the kinematics of the younger stars is possible (the internal dispersion in metallicities of dSph means that there is no need for all the accreted stars to have the same metallicities). These stars should all be on similar orbits, reflecting that of the parent satellite galaxy. All stars in the proper-motion selected sample of Carney et al. are observed at essentially the same spatial location, and thus the most obvious effect would be simply a small velocity dispersion, of order that of the parent galaxy. Space motions are available from Carney et al. There are 17 stars which contribute significantly to the expected value of 13.1, bluer than the 16 Gyr turnoff, quoted above in Section 2. They have $\sigma_{U,V,W} = (165, 126, 83)$ km/s, and $<V>_{lag} = 160$ km/s. The corresponding values for all the stars in this metallicity range are $\sigma_{U,V,W} = (157, 106, 72)$ km/s, and $<V>_{lag} = 173$ km/s. There is no obvious difference, strongly arguing against the hypothesis that the apparently younger stars were accreted from of order one satellite galaxy.

The luminosity function of the present retinue of dwarf galaxies in the Local Group is rather flat, as may be inferred from Fig. 5, together with the luminosity–metallicity relation. In contrast, hierarchical clustering scenarios, such as Cold Dark Matter dominated models, in general predict a rather steep mass function over this range of masses. The transformation from luminosity to mass is not that straightforward; the interpretation of dSph internal velocity dispersions, where available, in terms of dynamical mass, implies a smaller range in mass than in luminosity. [§] The conflict between theoretical predictions and observations remains, however, since both the distribution along the mass function and its slope are wrong. We address the question of whether merging/accretion could have modified the primordial mass spectrum to produce that of the Local Group dwarfs in another paper (Gilmore and Wyse, in prep).

The formation of the dominant old, metal-poor field halo population is outside the scope of this paper. However, we recall that low density, low mass, stellar and gaseous systems, are relatively easily tidally disrupted, or self-disrupted by feedback from internal star formation and supernovae. Thus a primordial population of such 'building-blocks' could long ago have been assimilated into the Galaxy, as part of the process of galaxy formation. The systems with higher internal binding energy would survive relatively more frequently, and the dSph galaxies discussed here could well be their descendants.

### 3.3 Inferences from Carbon Stars

An intermediate-age population also signals itself through its evolved stars, and in particular by luminous red stars. M giants are rare in the field halo, but a quantitative comparison with those seen in dSph is difficult. In addition to M giants, dSph frequently also contain a substantial population of carbon stars. Mould & Aaronson (1983) noted that the carbon stars traced an intermediate-age population in the Carina dwarf galaxy. Constraints on the contribution of such a population to the field halo requires large surveys for Carbon stars, which are only now becoming available. For example, Azzopardi & Lequeux (1992) find 50 spectroscopically confirmed normal C-stars in the Fornax dSph, from which they infer a total population of 77 C-stars, and they similarly deduce a total population of 11 C-stars in the Carina dSph. Surveys of the carbon star content of the halo are unfortunately rather incomplete, making the analysis uncertain. For example, Green et al. (1994) find a high-latitude faint carbon star count of 0.02 deg$^{-2}$, based on the presence of a single C-star in a field of size 52 deg$^2$. Not just small-number statistics confuse the issue in the halo, but also the identification of dwarf carbon stars, inferred to be perhaps as common as giant carbon stars (Green, Margon & MacConnell 1991; Warren et al. 1993).

---

[§] This is relevant for accretion/assimilation since dynamical friction acts to remove orbital energy from a satellite on a timescale that scales approximately inversely with the mass of the satellite (all other things being held constant).

If one were to extrapolate from this one carbon star, scaling 0.02 stars deg$^{-2}$ for a high-latitude field to a total halo value, using the Hernquist (1990) density profile for the halo, one predicts a halo C-star population of ∼3500. Using a luminosity for the halo of $M_V = -17.4$; and adopting $M_V = -13.2$ for Fornax, and $M_V = -9.2$ for Carina, gives normalised C-star/luminosity of unity for the halo, 1.2 for Fornax and 6.0 for Carina. This would suggest that the carbon star content is not a problem for scenarios with significant accretion of dSph into the halo. This conclusion is contrary to other analyses (e.g. van den Bergh 1994; here the most important source of the difference is the extrapolation to the number of carbon stars expected for the entire halo), and contrary to the much more robust constraints from blue stars presented earlier. This just illustrates that it is impossible at present to say anything reliable from carbon star statistics due to lack of data.

### 3.4 Chemical Elemental Abundance Signatures

The range of ages and chemical abundances within these systems indicate a level of self-enrichment, which implies some retention and/or re-capture of gas (e.g. Silk, Wyse and Shields 1987). This may related to the inference that the actual star formation rate, even in the active phase, is very low – a small fraction of a solar mass per year – which may be too low an energy injection rate for a coherent supernovae-driven wind to be initiated. Such low star formation rates are indicated at least for the Carina dwarf by the data of Smecker-Hane et al. (1995). Their data are further suggestive that the major star formation event in the Carina dwarf had a duration of several Gyr.

These low star formation rates, and possible re-cycling of gas through successive generations of stars, suggests that the bulk of the stars in dSph will show nucleosynthesis products of Type Ia supernova, which means relatively high iron content (e.g. review by Wheeler, Sneden & Truran 1989). Extremely low values of oxygen-to-iron may be achieved in closed systems which form stars after a hiatus in star formation, during which iron from Type Ia supernova continues to be ejected to the interstellar medium, without any accompanying oxygen. Such a star formation history has been inferred for the Magellanic Clouds, and indeed low oxygen-to-iron ratios have apparently resulted (see Gilmore & Wyse 1991).

This is to be contrasted with the three-times-solar value of the oxygen-to-iron ratio seen in halo field stars, generally attributed to enrichment by Type II supernovae alone (e.g. Wyse and Gilmore 1988; Nissen et al. 1994).

The samples of halo stars with reliable element ratios are not large, however. Specific observations of the element ratios in candidate 'young' halo stars have not been made with modern detectors, to be analysed with up-to-date stellar atmospheres. This will be an important observation, since many plausible star formation histories for dSph predict rather distinctive signatures – low oxygen-to-iron – if the young stars do result from accretion of intermediate age populations that formed in these systems.

## 4 IF NOT INTO THE STELLAR HALO, THEN TO WHERE?

The central regions of galaxies are obvious repositories of accreted systems, being the bottom of the local potential well, provided the accreted systems are sufficiently dense to survive disruption while sinking to the centre (e.g. Tremaine, Ostriker & Spitzer 1975). The mean metallicity of the bulge is now reasonably well-established at [Fe/H]∼ −0.3 dex (McWilliam & Rich 1994; Ibata & Gilmore 1995), with a significant spread below −1 dex, and above solar. Thus satellite galaxies that could have contributed significantly to the bulge are restricted to those of high metallicity, more like the Magellanic Clouds, or the most luminous dSph.

The Sagittarius dwarf spheroidal galaxy was discovered through spectroscopy of a sample of stars selected purely on the basis of colour and magnitude to contain predominantly K giants in the Galactic bulge. After rejection of foreground dwarf stars, the radial velocities isolated the Sagittarius dwarf galaxy member stars from the foreground bulge giants. The technique (serendipity) used to discover the Sagittarius dSph allows a real comparison between its stellar population and that of the bulge. Not only the radial velocities distinguish the dwarf galaxy, but also its stellar population – as seen in Fig. 1 of Ibata et al. (1994), all giant stars redder than $B_J - R \gtrsim 2.25$ have kinematics that place them in the low velocity-dispersion component *i.e.* in the Sagittarius dwarf. This is a real quantifiable difference between the *bulge* field population and this dwarf spheroidal galaxy.

As with the halo above, the carbon star population of the bulge can be compared with those of typical extant satellites. In this case there is a clear discrepancy between the bulge and the Magellanic Clouds and the dSph (Azzopardi & Lequeux 1992).

Although robust quantitative results are not yet possible on the age of the stellar population in the Galactic Bulge, it is substantially older than is the dominant stellar population of the Sagittarius dwarf, or of the Magellanic Clouds. The Galactic Bulge cannot have been created, at least in large part, from the accretion of a large number of dwarf satellites like these.

## 5 CONCLUSIONS

The age distribution, and chemical elemental abundances, of stars in the halo of the Milky Way provide constraints on theories of galaxy formation. As one specific example, the accretion of satellite galaxies similar to the present retinue of dwarf spheroidals (dSphs) would provide an observable metal-poor, intermediate-age population. This paper presents a quantitative assessment of any contribution made by such stars to the stellar halo. The bulk of the stellar populations in the halo show a well-defined turn-off, at B−V∼ 0.4, implying that the vast majority of the stars are old. The fraction of stars which lie blueward of this well-defined turn-off, with metallicities similar to that of the present dSphs, is used in this paper to place limits on the importance of the recent accretion of such systems. We adopt a conservative approach, leading to an upper limit of $\lesssim 10\%$ of halo stars being bluer (and by implication, younger) than this limit, with the highest value found for the more metal-rich halo ([Fe/H] $\gtrsim -1.5$ dex). Direct comparison of this statistic with the colour distribution of the turnoff stars in the Carina dwarf allows us to derive an upper limit on the number of mergers of such satellite galaxies into the halo of the Milky Way. This upper limit is $\sim 40$ Carina-like galaxies. The higher metallicity data constrain satellite galaxies like the Fornax dwarf; only $\lesssim 5$ of these could have been accreted. The approach taken to counting stars blueward of the dominant turnoff leads to a limit on not just recent accretion into the halo of systems with intermediate-age, metal-poor stars, but of accretion of such systems over the last $\gtrsim 10$ Gyr. Further, no satellite with the mass and metallicity of the SMC can have been accreted into the halo over this time. We note that the low star-formation rates inferred for dSph predict distinctive elemental abundance signatures; future data for field halo stars, including candidate younger stars, will provide a further robust test of accretion models.

The luminosity-weighted metallicity distribution of the present retinue of dSph galaxies is dominated by the most metal-rich systems, with [Fe/H] $\gtrsim -1$ dex. This contrasts strongly with the field halo. The more metal-rich central bulge of the Galaxy also could not have formed by accretion of systems with similar stellar populations to these metal-rich dwarfs, nor similar to the Magellanic Clouds. However, accretion is on-going, as evidenced by the Sagittarius dwarf, and the younger stellar population of each of the halo and bulge is likely to increase.

## 6 ACKNOWLEDGEMENTS

RFGW acknowledges support from the Seaver Foundation. Our collaboration is aided by the NSF (INT-9113306). The Center for Particle Astrophysics is supported by the NSF. MU acknowledges the financial support of the Particle Physics and Astronomy Research Council.

## REFERENCES

Aaronson M., 1986, in C. Norman, A. Renzini and M. Tosi, eds, Stellar Populations, Cambridge Univ. Press, p.45
Arnold R., Gilmore G., 1992, MNRAS, 257, 225
Armandroff T.E., DaCosta G.S, Caldwell N., Seiter P., AJ, 1993, 106, 986
Azzopardi M., Lequeux J., 1992, in Barbury B., Renzini A., eds, Proc. IAU Symp. 149, The Stellar Populations of Galaxies. Kluwer, Dordrecht, p.201
Baade W., 1944, ApJ, 100, 137
Beauchamp D., Hardy E., Suntzeff N.B., Zinn R., 1995, AJ, 109, 1628
Bok B.J., McRae D.A., 1941, Ann. New York Acad. of Sciences XLII, p.219
Bond H.E., MacConnell D.J. 1971, ApJ, 165, 51
Carney B.W., Latham D.W., Laird J.B., 1989, AJ, 97, 423
Carney B.W., Latham D.W., Laird J.B., Aguilar L.A., 1994, AJ, 107, 2240
Chaboyer B. et al., ApJ submitted
Conlon E.S., Dufton P.L., Keenan F.P., McCausland R.J.H., Holmgren D., 1992, ApJ, 400, 273
Da Costa G., Armandroff T., 1995, AJ, 109, 2533
Edvardsson B., Andersen J., Gustafsson B., Lambert D.L., Nissen P.E., Tomkin J., 1993, A&A, 275, 101
Elvius T., 1965, in Blaauw A., Schmidt M., eds, Stars & Stellar Systems Vol V – Galactic Structure, Chicago Univ. Press, p.41
Fusi Pecci F., Bellazzini M., Cacciari C., Ferraro F.R., 1995, AJ, in press (October)
Ferraro F.R., Pecci F.F., Bellazzini M., 1995, A&A, 298, 461
Gilmore G., Wyse R.F.G 1987 in Carswell R., Gilmore G., eds, The Galaxy, Reidel, Dordrecht, p.247
Gilmore G., Wyse R.F.G, 1991, ApJ, 367, L55
Gilmore G., Wyse R.F.G, Jones J.B., 1995, AJ, 109, 1095
Gilmore G., Wyse R.F.G., Kuijken K., 1989, ARA&A, 27, 555
Green E.M., Demarque P., King C.R., 1987. The Revised Yale Isochrones, Yale University Observatory.
Green P.J., Margon B., MacConnell D.J., 1991, ApJ, 380, L31
Green P.J., Margon B., Anderson S.F, Cook K.H., 1994, ApJ, 434, 319
Griffiths R.E. et al., 1994, ApJ, 435, L19
Hatzidimitriou D., Irwin M.J., 1995, in preparation
Hernquist L., 1990, ApJ, 356, 359
Ibata R.A., Gilmore G., 1995, MNRAS in press
Ibata R.A., Gilmore G., Irwin M.J., 1994, Nat, 370, 194
Laird J.B., Rupen M.P., Carney B.W., Latham D.W., 1988, AJ, 96, 1908
Lee M.G., Freedman W., Mateo M., Thompson I., Rath,. M., Ruiz, M.-T., 1993, AJ, 106, 1420
Lin D.N.C., Lynden-Bell D.L., 1977, MNRAS, 181, 59
Lindblad B., 1926, Medd. Fr. Astr. Obs. Upsala, Nova Acta Regiae Societatis Scientarum Upsaliensis
McWilliam A., Rich M., 1994, ApJS, 91, 749
Majewski, S. 1992, ApJS, 78, 87
Majewski S., 1993, in Majewski S., ed, ASP Conf. Series 49, Galaxy Evolution: The Milky Way Perspective, p.5
Marquez A., Schuster W.J., 1994, A&AS, 108, 341
Mateo M., Udalski A., Szymanski M., Kaluzny J., Kubiak M., Kreminski W., 1995, AJ, 109, 588
May A., Binney J., 1986, MNRAS, 221, 857
Mighell K.J., 1990, A&AS, 82, 1
Mould J., Aaronson M., 1983, ApJ, 273, 530
Nissen P.E., Gustafsson B., Edvardsson B., Gilmore G., 1994, A&A, 285, 440
Ostriker J.P., Tremaine S., 1975, ApJ, 202, L113
Preston G.W., Beers T.C., Shectman S.A., 1994, AJ, 108, 538
Reid N., Mould J., 1991, AJ, 101, 1299
Ryan S.G., Norris J.E., 1993, in Majewksi S.R., ed, ASP Conf. Series 49, Galaxy Evolution: The Milky Way Perspective, p.103
Quinn P.J., Hernquist L., Fullagar D.P., 1993, ApJ, 403, 74
Sandage A.R., 1969, ApJ, 158, 1115
Schuster W., Nissen P., 1989, A&A, 222, 89
Searle L., 1977, in Tinsley B.M., Larson R.B., eds, Evolution of Galaxies and Stellar Populations. Yale Univ. Obs., p.219
Silk J., Wyse R.F.G., 1993, Phys. Reports, 1993, 231, 293
Silk J., Wyse R.F.G., Shields G.A., 1987, ApJ, 322, L59
Smecker-Hane T.A., Stetson P.B., Hesser J.E., Lehnert M.D., 1994, AJ, 108, 507
Smecker-Hane T.A., Stetson P.B., Hesser J.E., 1995. In preparation.
Suntzeff N.B., Aaronson M., Olszweski E.W, Cook K.H., 1986, AJ, 91, 1091
Tinsley, B.M. and Gunn, J.E. 1976, ApJ, 203, 52
Toth G., Ostriker J.P., 1992, ApJ, 389, 5
Tremaine S.D., Ostriker J.P., Spitzer Jr, L., 1975, ApJ, 196, 407
Upgren A.R., 1963, AJ, 68, 475
VandenBerg D.A., 1985, ApJS, 58, 711
VandenBerg D.A., Bell R.A., 1985, ApJS, 58, 561
van den Bergh S., 1994, AJ, 108, 2145
Warren S.J., Irwin M.J., Evans D.W., Osmer P.S., Hewett P.C., 1993, MNRAS, 261, 185
Wheeler J.C., Sneden C., Truran J.W., 1989, ARA&A, 27, 279

Wyse R.F.G, Gilmore G., 1988, AJ, 95, 1404
Zinn R., 1993, in Smith G.H., Brodie J.P., eds, ASP conf. series 48, The Globular Cluster – Galaxy Connection, p.38